\documentclass[aps,prl,twocolumn,superscriptaddress,showpacs]{revtex4}
\usepackage{graphicx} 

\begin{document}
\lefthyphenmin=2
\righthyphenmin=3

%
%
\title{
Search for Exclusive {\boldmath $\gamma\gamma$} Production in Hadron-Hadron Collisions
}
\affiliation{Institute of Physics, Academia Sinica, Taipei, Taiwan 11529, Republic of China} 
\affiliation{Argonne National Laboratory, Argonne, Illinois 60439} 
\affiliation{Institut de Fisica d'Altes Energies, Universitat Autonoma de Barcelona, E-08193, Bellaterra (Barcelona), Spain} 
\affiliation{Baylor University, Waco, Texas  76798} 
\affiliation{Istituto Nazionale di Fisica Nucleare, University of Bologna, I-40127 Bologna, Italy} 
\affiliation{Brandeis University, Waltham, Massachusetts 02254} 
\affiliation{University of California, Davis, Davis, California  95616} 
\affiliation{University of California, Los Angeles, Los Angeles, California  90024} 
\affiliation{University of California, San Diego, La Jolla, California  92093} 
\affiliation{University of California, Santa Barbara, Santa Barbara, California 93106} 
\affiliation{Instituto de Fisica de Cantabria, CSIC-University of Cantabria, 39005 Santander, Spain} 
\affiliation{Carnegie Mellon University, Pittsburgh, PA  15213} 
\affiliation{Enrico Fermi Institute, University of Chicago, Chicago, Illinois 60637} 
\affiliation{Comenius University, 842 48 Bratislava, Slovakia; Institute of Experimental Physics, 040 01 Kosice, Slovakia} 
\affiliation{Joint Institute for Nuclear Research, RU-141980 Dubna, Russia} 
\affiliation{Duke University, Durham, North Carolina  27708} 
\affiliation{Fermi National Accelerator Laboratory, Batavia, Illinois 60510} 
\affiliation{University of Florida, Gainesville, Florida  32611} 
\affiliation{Laboratori Nazionali di Frascati, Istituto Nazionale di Fisica Nucleare, I-00044 Frascati, Italy} 
\affiliation{University of Geneva, CH-1211 Geneva 4, Switzerland} 
\affiliation{Glasgow University, Glasgow G12 8QQ, United Kingdom} 
\affiliation{Harvard University, Cambridge, Massachusetts 02138} 
\affiliation{Division of High Energy Physics, Department of Physics, University of Helsinki and Helsinki Institute of Physics, FIN-00014, Helsinki, Finland} 
\affiliation{University of Illinois, Urbana, Illinois 61801} 
\affiliation{The Johns Hopkins University, Baltimore, Maryland 21218} 
\affiliation{Institut f\"{u}r Experimentelle Kernphysik, Universit\"{a}t Karlsruhe, 76128 Karlsruhe, Germany} 
\affiliation{High Energy Accelerator Research Organization (KEK), Tsukuba, Ibaraki 305, Japan} 
\affiliation{Center for High Energy Physics: Kyungpook National University, Taegu 702-701, Korea; Seoul National University, Seoul 151-742, Korea; SungKyunKwan University, Suwon 440-746, Korea} 
\affiliation{Ernest Orlando Lawrence Berkeley National Laboratory, Berkeley, California 94720} 
\affiliation{University of Liverpool, Liverpool L69 7ZE, United Kingdom} 
\affiliation{University College London, London WC1E 6BT, United Kingdom} 
\affiliation{Centro de Investigaciones Energeticas Medioambientales y Tecnologicas, E-28040 Madrid, Spain} 
\affiliation{Massachusetts Institute of Technology, Cambridge, Massachusetts  02139} 
\affiliation{Institute of Particle Physics: University of Alberta, Edmonton, Canada T6G 2G7; McGill University, Montr\'{e}al, Canada H3A~2T8; and University of Toronto, Toronto, Canada M5S~1A7} 
\affiliation{University of Michigan, Ann Arbor, Michigan 48109} 
\affiliation{Michigan State University, East Lansing, Michigan  48824} 
\affiliation{University of New Mexico, Albuquerque, New Mexico 87131} 
\affiliation{Northwestern University, Evanston, Illinois  60208} 
\affiliation{The Ohio State University, Columbus, Ohio  43210} 
\affiliation{Okayama University, Okayama 700-8530, Japan} 
\affiliation{Osaka City University, Osaka 588, Japan} 
\affiliation{University of Oxford, Oxford OX1 3RH, United Kingdom} 
\affiliation{University of Padova, Istituto Nazionale di Fisica Nucleare, Sezione di Padova-Trento, I-35131 Padova, Italy} 
\affiliation{LPNHE, Universite Pierre et Marie Curie/IN2P3-CNRS, UMR7585, Paris, F-75252 France} 
\affiliation{University of Pennsylvania, Philadelphia, Pennsylvania 19104} 
\affiliation{Istituto Nazionale di Fisica Nucleare Pisa, Universities of Pisa, Siena and Scuola Normale Superiore, I-56127 Pisa, Italy} 
\affiliation{University of Pittsburgh, Pittsburgh, Pennsylvania 15260} 
\affiliation{Purdue University, West Lafayette, Indiana 47907} 
\affiliation{University of Rochester, Rochester, New York 14627} 
\affiliation{The Rockefeller University, New York, New York 10021} 
\affiliation{Istituto Nazionale di Fisica Nucleare, Sezione di Roma 1, University of Rome ``La Sapienza," I-00185 Roma, Italy} 
\affiliation{Rutgers University, Piscataway, New Jersey 08855} 
\affiliation{Texas A\&M University, College Station, Texas 77843} 
\affiliation{Istituto Nazionale di Fisica Nucleare, University of Trieste/\ Udine, Italy} 
\affiliation{University of Tsukuba, Tsukuba, Ibaraki 305, Japan} 
\affiliation{Tufts University, Medford, Massachusetts 02155} 
\affiliation{Waseda University, Tokyo 169, Japan} 
\affiliation{Wayne State University, Detroit, Michigan  48201} 
\affiliation{University of Wisconsin, Madison, Wisconsin 53706} 
\affiliation{Yale University, New Haven, Connecticut 06520} 
\author{T.~Aaltonen}
\affiliation{Division of High Energy Physics, Department of Physics, University of Helsinki and Helsinki Institute of Physics, FIN-00014, Helsinki, Finland}
\author{A.~Abulencia}
\affiliation{University of Illinois, Urbana, Illinois 61801}
\author{J.~Adelman}
\affiliation{Enrico Fermi Institute, University of Chicago, Chicago, Illinois 60637}
\author{T.~Affolder}
\affiliation{University of California, Santa Barbara, Santa Barbara, California 93106}
\author{T.~Akimoto}
\affiliation{University of Tsukuba, Tsukuba, Ibaraki 305, Japan}
\author{M.G.~Albrow}
\affiliation{Fermi National Accelerator Laboratory, Batavia, Illinois 60510}
\author{S.~Amerio}
\affiliation{University of Padova, Istituto Nazionale di Fisica Nucleare, Sezione di Padova-Trento, I-35131 Padova, Italy}
\author{D.~Amidei}
\affiliation{University of Michigan, Ann Arbor, Michigan 48109}
\author{A.~Anastassov}
\affiliation{Rutgers University, Piscataway, New Jersey 08855}
\author{K.~Anikeev}
\affiliation{Fermi National Accelerator Laboratory, Batavia, Illinois 60510}
% ...

%
\begin{abstract}
We have searched for exclusive $\gamma\gamma$ production in proton-antiproton 
collisions at 
$\sqrt{s}$ = 1.96 TeV, using 532 pb$^{-1}$ of integrated luminosity 
taken by the Run~II Collider Detector at Fermilab. 
The event signature requires two electromagnetic showers, each with transverse energy $E_T > 5$ GeV and
pseudorapidity $|\eta| <$ 1.0, with no other particles detected 
in the event. Three candidate events are observed.
We discuss the consistency of the three events with 
$\gamma\gamma, \pi^\circ\pi^\circ$, or $\eta\eta$ production. The probability that other processes
fluctuate to $\geq$ 3 events is $1.7 \times 10^{-4}$.
An upper limit on the cross section of $p\bar{p} \to p+\gamma\gamma+\bar{p}$ 
production is set at $410$~fb with 95\% confidence level.
\end{abstract} 

\maketitle

\begin{figure}[!ht]
  \begin{center}
    \includegraphics[width=0.50\textwidth]{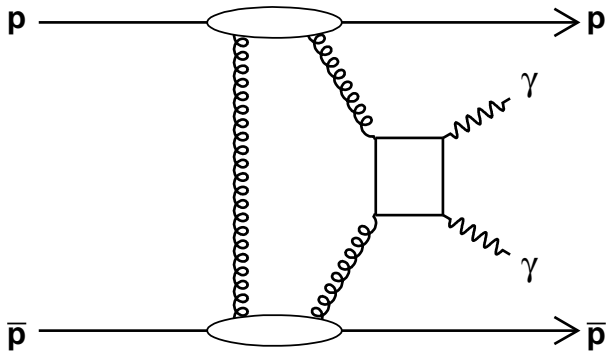}
    \caption{The dominant diagram for central exclusive $\gamma\gamma$ production 
    in $p\bar{p}$ collisions. The primary process is $gg\rightarrow\gamma\gamma$ through quark loops, with a
    screening gluon to cancel the exchanged color.\label{fig:exclusive_gg} }
  \end{center}
\end{figure}

We have searched for the ``exclusive" process $p+\bar{p} \rightarrow p+\gamma\gamma+\bar{p}$ in the Collider
Detector at Fermilab, CDF II, at $\sqrt{s}$ = 1.96 TeV. {\em Exclusive} means that no other particles are
produced; in our study the $p$ and $\bar{p}$ emerge intact with small transverse momenta $p_T$~\cite{kine} and the two photons
are central with pseudorapidity, $|\eta| <$ 1.0.
An exclusive $\gamma\gamma$ event can be 
produced via $gg \rightarrow \gamma \gamma$ through a quark loop, 
with an additional ``screening" gluon exchanged to cancel the color of the 
interacting gluons, and so allow the leading hadrons to stay intact, as shown in
Fig.~\ref{fig:exclusive_gg}. 
This process offers a novel
possibility to test QCD, and is closely related~\cite{cdfloi,KMRgg,exchiggs,forshaw} to exclusive Higgs boson~\cite{higgs} production at the LHC
$p+p\rightarrow p+H+p$, where the production mechanism 
of the Higgs boson is $gg$-fusion through a top quark loop. 
In both cases the final state,
$\gamma\gamma$ or $H$, is not strongly interacting, and thus the QCD calculation
of both diagrams is similar. However, the calculation is difficult
as the screening gluon has low $Q^2$, and other non-perturbative interactions in the same $p\bar{p}$ collision could 
produce additional particles. Calculations for exclusive Higgs 
production have been made using a variety of models, but these predictions cover a range of over two orders of 
magnitude~\cite{exchiggs,forshaw}. 
Since the QCD part of the calculation is the same for
$H$ and $\gamma\gamma$ production, and only the calculable matrix
elements $gg \to \gamma\gamma$ and $gg\to H$ 
are different, exclusive
$\gamma\gamma$ production provides an excellent test of the theoretical
predictions for $H$ production.  
For exclusive production
of two photons, each with transverse energy~\cite{kine} $E_T^\gamma>5$~GeV and pseudorapidity $|\eta^\gamma|<1$, the only predicted cross
sections~\cite{KMRgg} are $36$~fb at the Tevatron, at $\sqrt{s}$ = 1.96 TeV, and $200$~fb at the LHC.  
The same authors predict $\sigma(p+p \rightarrow p+H+p)$ = 3 fb at the LHC for a standard
model Higgs with $M_H = 120 - 140$ GeV/c$^2$, claiming a factor of about three uncertainty for both processes.
However, an NLO calculation has not been done and so these uncertainties are difficult to estimate.

Processes other than $gg \rightarrow \gamma\gamma$
can produce an exclusive $\gamma\gamma$ final state. 
Contributions from $q\bar{q} \rightarrow \gamma\gamma$ and
$\gamma\gamma \rightarrow \gamma\gamma$ are respectively $<$ 5\% and $<$ 1\% 
of $gg\rightarrow \gamma\gamma$ \cite{KMRgg}. The dominant backgrounds to 
the observation of exclusive $\gamma\gamma$ events are the production
of $\pi^\circ\pi^\circ$ or $\eta\eta$, with each meson decaying to two photons. 
No theoretical calculation of exclusive
$\pi^\circ\pi^\circ$ or $\eta\eta$ production has been published; however, both cross sections
are estimated~\cite{vkpriv} to be about 25\% of the diphoton process, in the kinematic range of this study. 

This Letter presents the first search for exclusive $\gamma\gamma$ production in hadronic interactions.
We use 532 pb$^{-1}$ integrated luminosity of $p\bar{p}$ collisions at
$\sqrt{s}=1.96$~TeV delivered to the CDF II detector at
the Tevatron.
The CDF II detector is a general purpose detector
described elsewhere~\cite{cdf};
here we give a brief summary of the detector components used in this
analysis. Surrounding the beam pipe is a tracking system consisting of
a silicon microstrip detector, a cylindrical drift
chamber (COT), and a solenoid providing a 1.4~Tesla
magnetic field.  The tracking system has nearly 100\% efficiency
for reconstructing isolated tracks with $p_T \geq$ 1 GeV/c and $|\eta|<1$.  
It is surrounded by the
central and end-plug calorimeters covering the
range $|\eta|<3.6$.  Both calorimeters have
separate electromagnetic and hadronic compartments. 
A proportional wire chamber (CES)~\cite{balka} is embedded in
the central electromagnetic calorimeter, $|\eta| < 1.1$, at a depth of six radiation lengths. It allows a measurement of the number and shape, in both transverse
directions, of electromagnetic showers. The anode wire pitch (in $\phi$) is 1.5~cm and 
the cathode strip pitch varies with $\eta$ from 1.7~cm to 2.0~cm. The CES provides a means of distinguishing
single photon showers from $\pi^\circ \rightarrow \gamma\gamma$ and $\eta \to\gamma\gamma$.
The region $3.6<|\eta|<5.2$ is covered by a
lead-liquid scintillator calorimeter called the
miniplug~\cite{miniplug}. At higher pseudorapidities,
$5.4<|\eta|<7.4$, scintillation counters, called beam shower
counters (BSC), are located on each side of the CDF detector.  Gas
$\check{\textrm{C}}$erenkov detectors covering 3.7 $<|\eta|<$ 4.7 determine the luminosity with a 6\%
uncertainty~\cite{clc}.

Exclusive $\gamma\gamma$ production is modeled with 
the \textsc{exhume}
Monte Carlo generator~\cite{exhume}, based on theoretical
calculations~\cite{KMRgg,KMR02}. 
Simulated single photons, and photons from $\pi^\circ$ and $\eta$ decay, are passed 
through the \textsc{geant}~\cite{geant}-based
detector simulation~\cite{cdfsim} to determine their detection efficiencies.

The event signature requires two 
electromagnetic showers each with transverse energy $E_T >$ 5 GeV, with no other particles detected in the full CDF detector, which
covers $-7.4 < \eta < +7.4$. The outgoing proton and antiproton are not detected. 
The event selection here follows closely that described in
Ref.~\cite{ee} where,
using the same trigger and a similar analysis, we observed exclusive $e^+e^-$ production. 
The only differences are the tracking requirements, and we
restrict the $|\eta|$ coverage from $\pm$2.0 to $\pm$1.0.
The trigger requires two electromagnetic clusters and
no BSC counter activity in the region $5.4 < |\eta| < 5.9$. The measured
cross section for $|\eta^e| < 2.0$ and $p_T^e \geq$~5~GeV/c 
(for both $e^+$ and $e^-$) is $1.6^{+0.5}_{-0.3}$(stat.)$\pm$ 0.3(syst.)~pb (16 candidates with 1.9$\pm$0.3 background), in agreement with the 
theoretical QED cross section of 1.71 $\pm$ 0.01 pb. Assuming the theoretical cross section to be correct, this agreement
is evidence that the efficiency of the cuts we make to define exclusive processes is well understood.

For the diphoton analysis we select events containing two electromagnetic showers, each
with $E_T>5$~GeV and $|\eta|<1.0$ and with 
a hadronic-to-electromagnetic energy ratio $< 0.058$, consistent with that of a photon. 
We require either no tracks pointing to the showers or two adjacent tracks consistent with a photon
conversion \mbox{($\gamma \rightarrow e^+e^-$)}.  
The efficiency for
triggering, reconstructing, and identifying a $\gamma\gamma$ event with two photons each with
\mbox{$E_T^\gamma>5$~ GeV} and \mbox{$|\eta^\gamma|<1$} is
\mbox{$0.57\pm 0.07$}.
Cosmic ray events are rejected
by requiring that the time of each shower is
consistent with photons coming from a bunch crossing. The efficiency of signal events to pass 
this cut is \mbox{$ 0.93 \pm 0.03$}. 
We define ``exclusivity" cuts that are designed to reject events having any additional particles
in the range $|\eta| <$ 7.4 that are not associated with the $\gamma$-candidates; these cuts require no
additional energy deposits (``particle signatures") above noise thresholds in the calorimeters or the BSC.
We do not use track or CES information in this selection.
One particle can shower and cause several ``signatures". 
We define the exclusivity cut efficiency \mbox{$\varepsilon_{\rm exc}$} as the probability that this exclusive requirement is not
spoiled by another inelastic interaction in the same bunch crossing. It is measured, as explained in Ref.~\cite{ee}, 
as the fraction of bunch crossing triggers that pass the exclusivity cuts,
which depends on the individual bunch-by-bunch luminosities. 
We find \mbox{$\varepsilon_{\rm exc}=0.086\pm 0.001$}.
The total efficiency is reduced by events which contain a photon conversion or electron bremsstrahlung
which fail the exclusivity requirements, estimated to be
\mbox{$0.87\pm0.09$} using the \textsc{exhume}
simulation. The probability of the scattered $p(\bar{p})$ depositing energy in the BSC is negligible
if their $p_T$ is less than 1.2 GeV/c.

The total efficiency for all the above event selection criteria, for the 
$p+\bar{p} \to p + \gamma\gamma + \bar{p}$ process, for photons each with $E_T^\gamma>5$~GeV and 
$|\eta^\gamma|<1$, is $4.0\pm0.7$\%. Three
events pass the selection criteria; their properties are given in Table I. The $E_T$ values of the six electromagnetic clusters are all between
5 and 7 GeV, and the azimuthal opening angle between the two photon candidates is 
$\Delta\phi_{\gamma\gamma} \geq 2.9$ rad, so the $\gamma\gamma$ invariant mass exceeds 10 GeV/c$^2$. 
The difference from $\Delta\phi_{\gamma\gamma} = \pi$ may be attributed to the outgoing
$p$ and $\bar{p}$ transverse momenta. 

\begin{table}[htbp]
\begin{center}
\begin{tabular}{cccccrrr}	
\hline
\hline
Event & S & $E_T$(GeV) & ($\eta$,$\phi$) & $N_{\rm CES}$ & $\chi^2_{\rm CES}$  & P($\pi^\circ$) & P($\gamma$) \\
\hline
A & A1 & 6.8 & (0.44,6.11) & 1 & 1.0 & 0.14 & 0.26 \\ 
  & A2 & 5.9 & (0.19,2.83) & 1 & 1.3 & 0.19 & 0.36  \\
B & B1 & 5.0 & (-0.07,4.86) & 1 & 1.4 & 0.21 & 0.39 \\
  & B2 & 5.4 & (0.67,1.66)  & 2 & -- & -- & -- \\
C & C1 & 6.0 & (-0.44,1.66) & 1 & 13.4 & 0.89 & 0.98 \\
  & C2 & 5.1 & (0.22,5.05)  & 2 & 2.2  & 0.33 & 0.57 \\
\hline
\hline
\end{tabular}
\end{center}
\caption{Properties of the calorimeter showers (S) of the three
candidate events: given are the $E_T$, the $\eta$ and $\phi$ location,
the total number of CES clusters inside the same CES chamber, $N_{\rm CES}$,
and the $\chi^2_{\rm CES}$ value (a shower shape variable, explained in the text).
Also given are the probabilities that a $\pi^\circ$ and a photon have a $\chi^2_{\rm CES}$ value smaller
than that observed.}
\label{tab:events} 
\end{table}

Five background sources to exclusive $\gamma\gamma$ production are considered:
cosmic rays, exclusive $e^+e^-$ events where both electrons are
mis-identified as photons, non-exclusive events in which 
additional particles do not leave a signature in the detector,
``quasi-exclusive" events where one or both outgoing protons dissociate
and the dissociation products are all very forward, beyond the detector coverage,
and exclusive $\pi^\circ\pi^\circ$ or $\eta\eta$ production.

The cosmic ray background is determined to be negligible
from the distribution of the arrival time of electromagnetic showers.
Cosmic rays are also expected to give hits in the tracking detectors.
However, a visual inspection of the event displays shows 
only random noise hits in
the COT and the silicon detector for 5 of the 6 showers in Table I. In the sixth case, shower B2,
an $e^+e^-$-pair from a photon conversion is seen, with the sum of the two
momenta consistent with the calorimeter shower energy.

Dielectron events~\cite{ee} could be misidentified as $\gamma\gamma$ events if both
electron tracks are not reconstructed or the electrons undergo energetic
bremsstrahlung.  This contribution is estimated from Ref.~\cite{ee} to be \mbox{$0.02\pm 0.02$} events.

Non-exclusive events, i.e. those with central particles in addition
to the two photons, may appear to be exclusive if the additional
particles are not detected through inefficiency.  
We study this by selecting events that contain two photon candidates
and no tracks in the tracking detectors (other than conversion tracks), 
without other requirements on the central and end-plug calorimeters. 
Only four events in the data
sample pass these criteria: the three candidates with zero
additional particle signatures, and one event with 13 signatures in the calorimeters.
This background is estimated to be $0.06 \pm 0.03$ events by using the same shape for the
distribution of additional particle signatures as in exclusive $e^+e^-$ events~\cite{ee}. 

The proton dissociation background is
small since all the dissociation
products must have $|\eta|>7.4$ to escape detection in the BSC counters. There are also
few excitation states available to the proton due to spin
restrictions on the final state~\cite{KMR02}. In Ref.~\cite{KMRgg} it was estimated that the
dissociation background is not expected to exceed $\sim 0.1\%$ of
the exclusive signal sample, which corresponds to 
$\leq 0.01$ events in the three-candidate sample. We take this background to be 
$0.01 \pm 0.01$ events.

Backgrounds can arise from exclusive pair 
production of neutral mesons, \textit{i.e.} 
$\pi^\circ\pi^\circ$ and $\eta\eta$. One photon from the $\pi^\circ$ or $\eta$ decay can be undetected or, in the
$\pi^\circ$ case, the two photon showers can merge. Exclusive $\pi^\circ\eta$ is suppressed by isospin conservation and $\gamma+(\pi^\circ/\eta)$ is forbidden
by C-parity conservation. 
Production of $\pi^\circ\pi^\circ$ and $\eta\eta$ cannot be unambiguously distinguished from $\gamma\gamma$ production on an event-by-event basis. 
Since the cross sections are not well known, these backgrounds cannot be directly calculated; we discuss them later. 

We therefore observe three exclusive $p+\bar{p}\rightarrow p+(\gamma\gamma/\pi^\circ\pi^\circ/\eta\eta)+ \bar{p}$ candidate events with a background of 
$0.09 \pm 0.04$ events. The probability for three or more events to be observed when 0.09 $\pm$ 0.04
(assumed to be the mean and standard deviation of a gamma-distribution) are expected is 1.7$\times 10^{-4}$.
We set an upper limit on the cross section for exclusive $\gamma\gamma$ production, taking into account
the background and its uncertainty, the signal selection efficiency, 
and the integrated luminosity. A Bayesian approach is used assuming a flat prior for the 
cross section and a gamma-distribution for the
uncertainties. 
This gives a limit on the production cross section 
$\sigma \left(p+\bar{p}\rightarrow p+\gamma\gamma+\bar{p}\right) <$ 410 fb (for
$E_T^\gamma>5\mbox{ GeV}, |\eta^\gamma|<1$) at 95\% confidence level.

We now discuss the three candidate events as possible $\gamma\gamma$, $\pi^\circ\pi^\circ$, or $\eta\eta$
production. 
The selection efficiency, including exclusivity cuts, for a photon 
from an isolated $\pi^\circ\rightarrow \gamma\gamma$ is 13\% lower than that of a direct photon, while the selection 
efficiency for an isolated $\eta\rightarrow \gamma\gamma$ is 35\% lower. 
Relative to $\pi^\circ\pi^\circ$ production, $\eta\eta$ detection is further suppressed by 
a factor 0.15, due to the branching fraction for $\eta\rightarrow\gamma\gamma$.
We therefore treat the potential background as being predominantly $\pi^\circ\pi^\circ$.

\begin{figure}[htbp]
  \begin{center}
    \includegraphics[width=0.5\textwidth]{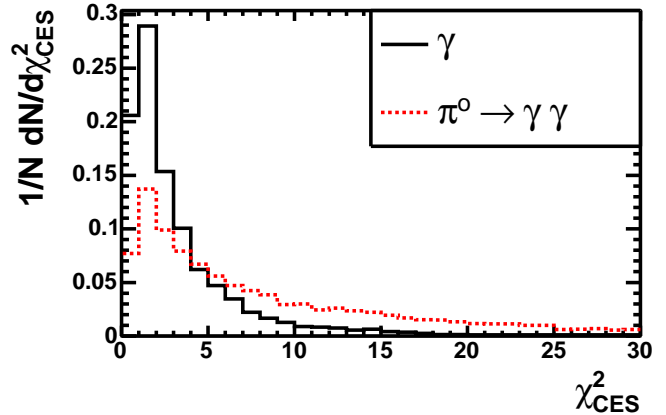}
    \caption{The simulated distribution for $\chi^2_{\rm CES}$ for prompt photons (solid histogram), 
and $\pi^\circ \to\gamma\gamma$ decays (dashed histogram). 
In all cases $E_T$ is required to be between $5$ and $7$~GeV.
\label{fig:chi2} }
  \end{center}
\end{figure}

We can only distinguish between single photons, and photons from $\pi^\circ$ decay, using the distribution
of signals on the CES strips and wires, in the module covering $\Delta\phi = 15^\circ$ and $|\eta|< 1.1$
which contains the shower. 
CES clusters are formed using 11 adjacent strips or wires. We may observe two separate clusters, $N_{\rm CES}$ = 2,
from $\pi^\circ$ decay. If we observe a single cluster, $N_{\rm CES}$ = 1, it could be from a $\pi^\circ$ if
the two photon showers overlap or if one photon shower is not detected.  
The number of CES clusters in the three candidate events is shown in Table I.
While only 12\% of photons have a 
second CES cluster, 28\% (46\%) of the $\pi^\circ$ ($\eta$) do. 
From simulation the probability that one photon from $\pi^\circ/\eta \rightarrow \gamma\gamma$ is not detected in
the CES, by ranging out or not interacting, is 0.125$\pm$0.025. Single clusters from photons or $\pi^\circ$ can be distinguished statistically
using their shape. We use the distribution of pulse heights on the wires and strips to form a variable,  $\chi^2_{\rm CES}$, that
compares the lateral shape with that for an electron shower. 
A simulated distribution of $\chi^2_{\rm CES}$
for photons and $\pi^\circ$'s is shown in Fig.~\ref{fig:chi2}; it has a longer tail
for $\pi^\circ$'s than for photons but it does not allow an event-by-event
separation. 
Using the distributions in~Fig.~\ref{fig:chi2} the probability ($P(\gamma),P(\pi^\circ)$) that 
a shower has a $\chi^2_{\rm CES}$
less than the observed value was calculated for the five non-conversion shower candidates. Calculated values are given in
Table I.

In Event A both showers are single clusters with a small $\chi^2_{CES}$, more consistent with 
originating from photons than from $\pi^\circ$'s.
In Event B shower B1 also has a very low $\chi^2_{CES}$, while shower B2 is a photon conversion and
the $\chi^2_{\rm CES}$ method cannot be
used. Two clusters in the CES are separated in $\phi$, but not in $\eta$, as expected for a conversion. 
The sum of the two track momenta is 5.40 GeV/c, and the calorimeter energy is $5.45 \pm 0.35$ GeV, so 
if there were a second photon from
a $\pi^\circ$ or $\eta$ it would have $E_\gamma <$ 0.55 GeV (95\% C.L.), with a probability $<10$\% that a $\pi^\circ$ or $\eta$ 
decay would have such an energy asymmetry. Also, no additional shower is observed.
Therefore events A and B clearly favor the $\gamma\gamma$
hypothesis, with three narrow single showers and one photon conversion without an accompanying shower. 
We cannot give an unbiased value for the $\pi^\circ\pi^\circ$ background,
since this study was done {\em a posteriori}.
In Event C, one shower (C1) has a very large $\chi^2_{\rm CES}$ (only 2\% of photon showers have a larger value), 
and the other shower (C2) has
$N_{\rm CES}$ = 2. Both favor the hypothesis that it is a $\pi^\circ\pi^\circ$ event. A likelihood ratio calculation, using only
the $N_{\rm CES}$ and $\chi^2_{\rm CES}$ distributions, favors the $\pi^\circ\pi^\circ$ hypothesis
over the $\gamma\gamma$ hypothesis by a factor 4.
For one event in three candidates to be $\pi^\circ\pi^\circ$ is
compatible with the theoretical estimate~\cite{vkpriv} (about 1/5).  


In conclusion, we have observed three candidate events for exclusive $\gamma\gamma$,
$\pi^\circ\pi^\circ$, or $\eta\eta$ production with an expected background of $0.09 \pm 0.04$ events. 
The probability to observe three or more events when $0.09 \pm 0.04$ are expected from other processes is
$1.7 \times 10^{-4}$, corresponding to a statistical significance of $3.7\sigma$. Though two of the candidates 
are most likely to arise from 
$\gamma\gamma$ production, the $\pi^\circ\pi^\circ/\eta\eta$ hypotheses cannot be excluded. Therefore we report a 95\% C.L. upper limit on the exclusive $\gamma\gamma$
production cross section ($E_T^\gamma > 5$ GeV, $|\eta^\gamma| < $ 1.0) of 410~fb, approximately a factor of ten higher 
than the prediction~\cite{KMRgg}. 
We note that the prediction of Ref.~\cite{KMRgg} of $36^{+72}_{-24}$ fb
would correspond to $0.8^{+1.6}_{-0.5}$ events, compatible with our observations.
This result may be used to constrain calculations of exclusive Higgs boson production at the LHC; it
disfavors the highest predictions. 

We thank the Fermilab staff and the technical staffs
of the participating
institutions for their vital contributions. This
work was supported by the
U.S. Department of Energy and National Science
Foundation; the Italian Istituto
Nazionale di Fisica Nucleare; the Ministry of
Education, Culture, Sports,
Science and Technology of Japan; the Natural
Sciences and Engineering Research
Council of Canada; the National Science Council of
the Republic of China; the
Swiss National Science Foundation; the A.P. Sloan
Foundation; the
Bundesministerium f\"ur Bildung und Forschung,
Germany; the Korean Science and
Engineering Foundation and the Korean Research
Foundation; the Science and Technology Facilities Council and the Royal
Society, UK; the Institut National de Physique Nucleaire et Physique des Particules/CNRS, the Russian Foundation
for Basic Research; the Comisi\'on Interministerial
de Ciencia y
Tecnolog\'{\i}a, Spain; the European Community's Human Potential Programme;
the Slovak R\&D Agency, and the
Academy of Finland.


\begin{thebibliography}{99}
\bibitem{kine} A cylindrical coordinate system is used with the $z$-axis along the 
proton beam direction; $\theta$ is the polar angle and $\phi$ is the azimuthal angle.
We define pseudorapidity as $\eta= -\ln \tan(\theta/2)$,
transverse momentum as 
$p_T =|p| \sin\theta$, and transverse energy as $E_T = E \sin\theta$.
\end{thebibliography}
\end{document}